\documentclass{article}
\usepackage{amsmath,amssymb}
\numberwithin{equation}{section}
\newcommand{\beq}{\begin{eqnarray}}
\newcommand{\eeq}{\end{eqnarray}}
\newcommand{\beqnn}{\begin{eqnarray*}}
\newcommand{\eeqnn}{\end{eqnarray*}}
\newcommand{\rd}{\partial}

\newcommand{\CC}{\mathbf{C}}
\newcommand{\PP}{\mathbf{P}}

\newcommand{\ZZ}{\mathbf{Z}}
\newcommand{\bsa}{\boldsymbol{a}}
\newcommand{\bsb}{\boldsymbol{b}}
\newcommand{\bst}{\boldsymbol{t}}

\newcommand{\bstbar}{\bar{\bst}}
\newcommand{\bsx}{\boldsymbol{x}}

\newcommand{\tbar}{\bar{t}}
\newcommand{\ubar}{\bar{u}}

\newcommand{\wbar}{\bar{w}}
\newcommand{\Vbar}{\bar{V}}
\newcommand{\Wbar}{\bar{W}}
\newcommand{\Bbar}{\bar{B}}
\newcommand{\Lbar}{\bar{L}}
\newcommand{\Psibar}{\bar{\Psi}}
\newcommand{\calL}{\mathcal{L}}
\newcommand{\Nbar}{\bar{N}}

\begin{document}

\title{Two extensions of 1D Toda hierarchy}
\author{Kanehisa Takasaki\\
Graduate School of Human and Environmental Studies\\
Kyoto University\\
Yoshida, Sakyo, Kyoto, 606-8501, Japan\\
takasaki@math.h.kyoto-u.ac.jp}
\date{}
\maketitle

\begin{abstract}
The extended Toda hierarchy of Carlet, Dubrovin and Zhang 
is reconsidered in the light of a $2+1$D extension 
of the 1D Toda hierarchy constructed by Ogawa.  These two 
extensions of the 1D Toda hierarchy turn out to have 
a very similar structure, and the former may be thought of
as a kind of dimensional reduction of the latter. In particular, 
this explains an origin of the mysterious structure 
of the bilinear formalism proposed by Milanov.  
\end{abstract}

\begin{flushleft}
\end{flushleft}

\section{Introduction}

Geometry of 2D topological field theories has 
a profound relationship with integrable hierarchies 
\cite{Dubrovin-geom2dtft}.  Of particular interest 
is the topological sigma model (geometrically, 
the Gromov-Witten invariants) of the Riemann sphere 
$\CC\PP^1$, which is related to the 1D Toda hierarchy 
and its dispersionless limit.  To describe 
the correlation functions of ``descendants'' 
of primary observables, however, one has to extend 
the usual 1D Toda hierarchy by an extra set 
of commuting flows \cite{EY94,Getzler01,Zhang02}.   
In the following, we refer to this extension 
as ``logarithmic'', because the Lax equations of 
these extra commuting flows are formulated 
by a kind of logarithm of the Lax operator. 
Carlet, Dubrovin and Zhang presented 
a rigorous formulation of the logarithm of 
the Lax operator, and thereby formulated a Lax formalism 
of the extended Toda hierarchy \cite{CDZ04}. 

Recently, Milanov presented a bilinear (or Hirota) formalism 
of this extended Toda hierarchy \cite{Milanov05}. 
According to Milanov's results, the tau function 
of the usual 1D Toda hierarchy can be extended 
to this hierarchy and satisfies a bilinear equation. 
This equation is certainly an extension 
of the familiar bilinear equation 
(of the contour integral type \cite{JM83}) 
of the 1D Toda hierarchy, reducing to the latter 
as some of arbitrary constants 
in the equation are set to $0$.  For nonzero values 
of those arbitrary constants, however, Milanov's 
bilinear equation takes a quite mysterious form, 
the meaning of which has remained to be elucidated. 

In this paper, we propose to understand 
the logarithmic extension in the light of 
the so called ``$2+1$D extension''.  
Here ``1'' means the (lowest) temporal dimension, 
and ``2'' an extension of the spatial dimension 
(in the case of the Toda hierarchy, a 1D lattice) 
by an extra spatial dimension.  For example, 
a $2+1$D extension of the KdV equation 
was introduced by Calogero \cite{Calogero75},
Bogoyavlensky \cite{Bogoyavlenski90} and 
Schiff \cite{Schiff92} in different contexts.  
A Lie algebraic interpretation of the same equation 
and the associated hierarchy of commuting flows 
was discovered later \cite{Billig99,ISW99,IT01} 
and generalized to the nonlinear Schr\"odinger 
hierarchy \cite{KIT02}.  An important outcome 
of the Lie-algebraic studies is a systematic derivation 
of a bilinear formalism of those $2+1$D extensions.   
As regards the 1D Toda hierarchy, 
two different $2+1$D extensions (based on  
two different reductions of the 2D Toda hierarchy 
\cite{UT84} to the 1D Toda hierarchy) 
were constructed by Ogawa \cite{Ogawa08}.  
We shall show that the logarithmic extension 
of the 1D Toda hierarchy can be rewritten to a form 
that resembles one of Ogawa's $2+1$D extension. 
This enables us to consider the logarithmic extension 
as a kind of ``dimensional reduction'' 
of the $2+1$D extension.  We can thus derive 
a bilinear formalism of the logarithmic extension 
by the same method as used for the $2+1$D extensions 
\cite{IT01,KIT02,Ogawa08}.  Milanov's results 
can be thus recovered from the perspectives 
of $2+1$D extensions.  

This paper is organized as follows.  
Section 2 is a review of the Lax formalism 
and the bilinear formalism of the 1D Toda hierarchy.  
Since the 1D Toda hierarchy can be derived 
from the 2D Toda hierarchy, we omit the proof 
of the existence of the dressing operators 
and the tau function (which is parallel to the case 
of the 2D Toda hierarchy) and explain 
the derivation of bilinear equations in detail. 
Section 3 is a review of one of Ogawa's $2+1$D 
extensions that is relevant to the subject of this paper.  
Since Ogawa's paper \cite{Ogawa08} is rather sketchy 
on this case, we give a rather detailed account 
of its Lax and bilinear formalisms.  
Armed with the knowledge on the $2+1$D extension, 
we turn to Carlet, Dubrovin and Zhang's 
logarithmic extension in Section 4.  
We conclude this paper with a few remarks in Section 5.

\section{1D Toda hierarchy}

\subsection{Lax equations}

Let $s$ denote the spatial coordinate of the 1D Toda hierarchy. 
Unlike the usual formulation on a 1D lattice, 
$s$ is now understood to be a continuous variable.  
The Lax operator of the 1D Toda hierarchy 
is a difference operator of the form 
\beqnn
  \calL = e^{\rd_s} + b(s) + c(s)e^{-\rd_s}, 
\eeqnn
where $e^{n\rd_s}$'s ($\rd_s = \rd/\rd_s$) denote 
the shift operators that act on a function of $s$ 
as $e^{n\rd_s}f(s) = f(s+n)$, and $b(s)$ and $c(s)$ 
are dynamical variables. 
Time evolutions $\calL = \calL(\bst)$, 
$\bst = (t_1,t_2,\ldots)$, of the Lax operator 
are defined by the Lax equations 
\beq
  \frac{\rd\calL}{\rd t_n} = [A_n,\calL], 
  \quad n = 1,2,\ldots. 
\label{Lax-eq}
\eeq
The generators $A_n$ of time evolutions 
are constructed from $\calL$ as 
\beqnn
  A_n = \frac{1}{2}\left(\calL^n\right)_{\ge 0} 
      - \frac{1}{2}\left(\calL^n\right)_{<0}, 
\eeqnn
where $(\quad)_{\ge 0}$ and $(\quad)_{<0}$ denote 
the nonnegative and negative power parts 
of difference operators: 
\beqnn
  \left(\sum_{n=-\infty}^\infty a_n(s)e^{n\rd_s}\right)_{\ge 0} 
  = \sum_{n \ge 0}a_n(s)e^{n\rd_s}, \\
  \left(\sum_{n=-\infty}^\infty a_n(s)e^{n\rd_s}\right)_{<0} 
  = \sum_{n<0}a_n(s)e^{n\rd_s}. 
\eeqnn
The lowest ($n = 1$) Lax equation consists of 
the equations 
\beqnn
  \frac{\rd b(s)}{\rd t_1} = c(s+1) - c(s), \quad 
  \frac{\rd c(s)}{\rd t_1} = c(s)(b(s) - b(s-1)), 
\eeqnn
which can be converted to the usual 1D Toda equation 
\beqnn
  \frac{\rd^2\phi(s)}{\rd t_1^2} 
  = e^{\phi(s)-\phi(s+1)} - e^{\phi(s-1)-\phi(s)}
\eeqnn
by the change of variables 
\beqnn
  b(s) = \frac{\rd\phi(s)}{\rd t_1}, \quad 
  c(s) = e^{\phi(s-1)-\phi(s)}. 
\eeqnn

\subsection{Wave functions and auxiliary linear equations}

Let $W$ and $\Wbar$ be dressing operators of the form 
\beqnn
  W = 1 + \sum_{n=1}^\infty w_n(s)e^{-n\rd_s},\quad 
  \Wbar = \sum_{n=0}^\infty \wbar_n(s)e^{n\rd_s} 
\eeqnn
by which the Lax operator is expressed as 
\beq
  \calL = We^{\rd_s}W^{-1} = \Wbar e^{-\rd_s}\Wbar^{-1}. 
\label{L-WWbar}
\eeq
One can tune these dressing operators 
to satisfy the evolution equations 
\beq
  \frac{\rd W}{\rd t_n} = A_nW - \frac{1}{2}We^{n\rd_s}, \quad 
  \frac{\rd \Wbar}{\rd t_n} = A_n\Wbar + \frac{1}{2}\Wbar e^{-n\rd_s} 
\label{Sato-eq}
\eeq
as well.  These equations can be converted 
to the auxiliary linear equations 
\beq
  \frac{\rd\Psi(s,z)}{\rd t_n} = A_n\Psi(s,z), \quad 
  \frac{\rd\Psibar(s,z)}{\rd t_n} = A_n\Psibar(s,z)
\label{PsiA-lineq}
\eeq
for the wave functions 
\beqnn
  \Psi(s,z) = Wz^se^{\xi(\bst,z)/2} 
       = \left(1 + \sum_{n=1}^\infty w_n(s)z^{-n}
         \right)z^se^{\xi(\bst,z)/2}, \\
  \Psibar(s,z) = \Wbar z^se^{-\xi(\bst,z^{-1})/2}
       = \left(\sum_{n=0}^\infty \wbar_n(s)z^n
         \right)z^se^{-\xi(\bst,z^{-1})/2}, 
\eeqnn
where 
\beqnn
  \xi(\bst,z) = \sum_{n=1}^\infty t_nz^n. 
\eeqnn
The dressing relations (\ref{L-WWbar}), too, 
become auxiliary linear equations of the form 
\beq
  \calL\Psi(s,z) = z\Psi(s,z), \quad 
  \calL\Psibar(s,z) = z^{-1}\Psibar(s,z). 
\label{PsiL-lineq}
\eeq

\subsection{Bilinear equations for wave functions}

Let us introduce the difference operators 
\beqnn
  V = e^{-\rd_s}(W^*)^{-1}e^{\rd_s},\quad 
  \Vbar = e^{-\rd_s}(V^*)^{-1}e^{\rd_s}, 
\eeqnn
where $A^*$ denotes the formal adjoint of 
a difference operator $A$, namely, 
\beqnn
  \left(\sum_{n=-\infty}^\infty a_n(s)e^{n\rd_s}\right)^* 
  = \sum_{n=-\infty}^\infty e^{-n\rd_s}a_n(s), 
\eeqnn
and define the ``dual wave functions'' as 
\beqnn
  \Psi^*(s,z) = Vz^{-s}e^{-\xi(\bst,z)/2},\quad 
  \Psibar^*(s,z) = \Vbar z^{-s}e^{\xi(\bst,z^{-1})/2}. 
\eeqnn
As we show below, the wave functions $\Psi(s,z),\Psibar(s,z)$ 
and their duals satisfy the bilinear equation 
\beq
  \oint\frac{dz}{2\pi i}z^k\Psi(s',\bst',z)\Psi^*(s,\bst,z) 
= \oint\frac{dz}{2\pi i}z^{-k}\Psibar(s',\bst',z)\Psibar^*(s,\bst,z)
\label{Psi-bilineq}
\eeq
for $k = 0,1,2,\ldots$ and arbitrary values 
of $s',s,\bst',\bst$ except for the condition 
\footnote{If the spatial variable $s$ is integer-valued, 
this condition is obviously satisfied. 
Since $s$ is now a continuous variable, 
this condition is necessary to ensure single-valuedness 
of the integrands in (\ref{Psi-bilineq}).}
\beq
  s' - s \in \ZZ. 
\label{s's-condition}
\eeq
In the present setting, both hand sides 
of the bilinear equation may be thought of 
as the residue of formal Laurent series, namely, 
\beqnn
  \oint \frac{dz}{2\pi i}\sum_{n=-\infty}^\infty a_nz^n = a_{-1}, 
\eeqnn
though, in a complex analytic setting, 
they are understood to be the contour integrals 
along simple closed curves $C_\infty, C_0$ 
encircling the points $z = \infty,0$. 

A technical clue to the derivation of (\ref{Psi-bilineq}) 
is the the identity (see, e.g., Ogawa's paper \cite{Ogawa08})
\beq
  \oint\frac{dz}{2\pi i}\psi(s,z)\phi^*(s,z) 
  = (Ae^{\rd_s}B^*)_{s's} 
  = (Be^{-\rd_s}A^*)_{ss'} 
\label{key-formula}
\eeq
that holds, under condition (\ref{s's-condition}), 
for any difference operators 
\beqnn
  A = \sum_{n=-\infty}^\infty a_n(s)e^{n\rd_s},\quad 
  B = \sum_{n=-\infty}^\infty  b_n(s)e^{n\rd_s}
\eeqnn
and the associated ``wave functions'' 
\beqnn 
  \psi(s,z) = Az^s = \sum_{n=-\infty}^\infty a_n(s)z^{n+s},\quad
  \phi^*(s,z) = Bz^{-s} = \sum_{n=-\infty}^\infty b_n(s)z^{-n-s}. 
\eeqnn
$(\quad)_{s's}$ denotes the ``$(s',s)$-matrix element''
\footnote{If the spatial variable $s$ is integer-valued, 
this is indeed the matrix element of a $\ZZ\times\ZZ$ 
matrix that represents the difference operator.}
of difference operators: 
\beqnn
  \left(\sum_{n=-\infty}^\infty a_n(s)e^{n\rd_s}\right)_{s's} 
  = a_{s-s'}(s'). 
\eeqnn

We apply this formula to the operator relation 
\beqnn
  We^{\rd_s}V^* = e^{\rd_s} = \Wbar e^{\rd_s}\Wbar^* 
\eeqnn
and obtain the bilinear equation 
\beq
  \oint\frac{dz}{2\pi i}\Psi(s',\bst,z)\Psi^*(s,\bst,z) 
= \oint\frac{dz}{2\pi i}\Psibar(s',\bst,z)\Psibar^*(s,\bst,z), 
\label{Psi-bilineq0}
\eeq
which is a special case of (\ref{Psi-bilineq}) 
where $k = 0$ and $\bst' = \bst$.   
We can deform this equation to (\ref{Psi-bilineq}) 
by two steps as follows.  

The first step is to insert $z^{\pm k}$, $k = 0,1,2,\ldots$ 
into the contour integrals.  To this end, 
we apply $\calL^k$ to both hand sides of (\ref{Psi-bilineq0}) 
with respect to the variable $s'$ as 
\beqnn
  \oint\frac{dz}{2\pi i}\calL^k\Psi(s',\bst,z)\cdot\Psi^*(s,\bst,z) 
= \oint\frac{dz}{2\pi i}\calL^k\Psibar(s',\bst,z)\cdot\Psibar^*(s,\bst,z). 
\eeqnn
By (\ref{PsiL-lineq}), this equation turns into the equation 
\beq
  \oint\frac{dz}{2\pi i}z^k\Psi(s',\bst,z)\Psi^*(s,\bst,z) 
= \oint\frac{dz}{2\pi i}z^{-k}\Psibar(s',\bst,z)\Psibar^*(s,\bst,z). 
\label{Psi-bilineq1}
\eeq

The second step is to shift the value of $\bst$ 
in $\Psi(s',\bst,z)$ and $\Psibar(s',\bst,z)$.  
To this end, let us note that the auxiliary linear equations 
(\ref{PsiA-lineq}) can be extended to higher orders as 
\beqnn
\begin{aligned}
  \prod_{i=1}^\infty\left(\frac{\rd}{\rd t_i}\right)^{l_i}\Psi(s,z) 
  &= A_{l_1,l_2,\ldots}\Psi(s,z),\\
  \prod_{i=1}^\infty\left(\frac{\rd}{\rd t_i}\right)^{l_i}\Psibar(s,z) 
  &= A_{l_1,l_2,\ldots}\Psibar(s,z), 
\end{aligned}
\eeqnn
where $A_{l_1,l_2,\ldots}$'s are difference operators 
of finite order in $s$ that are recursively determined 
by $A_n$'s.  For example, 
\beqnn
  \frac{\rd^2\Psi(s,z)}{\rd t_m\rd t_n} 
  = \frac{\rd}{\rd t_m}\left(\frac{\rd\Psi(s,z)}{\rd t_n}\right) 
  = \left(\frac{\rd A_n}{\rd t_m} + A_nA_m\right)\Psi(s,z), 
\eeqnn
hence 
\beqnn
  A_{m,n} =  \frac{\rd A_n}{\rd t_m} + A_nA_m. 
\eeqnn
The same equation holds for $\Psibar(s,z)$ as well.  
Applying $A_{l_1,l_2,\ldots}$ to both hand sides 
of (\ref{Psi-bilineq1}) with respect to $s'$, 
we have  the equations 
\begin{multline*}
  \oint\frac{dz}{2\pi i} 
  z^k\prod_{i=1}^\infty
     \left(\frac{\rd}{\rd t_i}\right)^{l_i}\Psi(s',\bst,z)
     \cdot\Psi^*(s,\bst,z) \\
= \oint\frac{dz}{2\pi i}
  z^{-k}\prod_{i=1}^\infty
     \left(\frac{\rd}{\rd t_i}\right)^{l_i}\Psibar(s',\bst,z)
     \cdot\Psibar^*(s,\bst,z) 
\end{multline*}
for all values of $l_1,l_2,\ldots$.  
Since the derivatives of $\Psi(s',\bst,z)$ 
and $\Psibar(s',\bst,z)$ can be collected 
to the generating functions 
\beqnn
\begin{aligned}
  \sum_{l_1,l_2,\ldots=0}^\infty 
  \prod_{i=1}^\infty \frac{a_i^{l_i}}{l_i!}
  \left(\frac{\rd}{\rd t_i}\right)^{l_i}\Psi(s',\bst,z) 
    &= \Psi(s',\bst+\bsa,z),\\
  \sum_{l_1,l_2,\ldots=0}^\infty 
  \prod_{i=1}^\infty \frac{a_i^{l_i}}{l_i!}
  \left(\frac{\rd}{\rd t_i}\right)^{l_i}\Psibar(s',\bst,z) 
    &= \Psibar(s',\bst+\bsa,z) 
\end{aligned}
\eeqnn
of new variables $\bsa = (a_1,a_2,\ldots)$, 
the last bilinear equations can be converted 
to the generating functional form 
\begin{multline}
  \oint\frac{dz}{2\pi i}
  z^k\Psi(s',\bst+\bsa,z)\cdot\Psi^*(s,\bst,z)\\
= \oint\frac{dz}{2\pi i}
  z^{-k}\Psibar(s',\bst+\bsa,z)\cdot\Psibar^*(s,\bst,z). 
\label{Psi-bilineq2}
\end{multline}
Replacing $\bst+\bsa \to \bst'$, we obtain 
the bilinear equation (\ref{Psi-bilineq}). 

Though we omit details, one can conversely derive 
the auxiliary linear equations (\ref{PsiA-lineq}) 
and (\ref{PsiL-lineq}) from (\ref{Psi-bilineq}).

\subsection{Tau function and bilinear equations}

The wave functions and their duals can be expressed 
in terms of the tau function $\tau(s,\bst)$ as 
\beq
\begin{aligned}
  \Psi(s,z) 
&= \frac{\tau(s,\bst-[z^{-1}])}{\tau(s,\bst)}z^se^{\xi(\bst,z)/2},\\
  \Psi^*(s,z) 
&= \frac{\tau(s,\bst+[z^{-1}])}{\tau(s,\bst)}z^{-s}e^{-\xi(\bst,z)/2},\\
  \Psibar(s,z)
&= \frac{\tau(s+1,\bst+[z])}{\tau(s,\bst)}z^se^{-\xi(\bst,z^{-1})/2},\\
  \Psibar^*(s,z)
&= \frac{\tau(s-1,\bst-[z])}{\tau(s,\bst)}z^{-s}e^{\xi(\bst,z^{-1})/2}, 
\end{aligned}
\label{Psi-tau}
\eeq
where 
\beqnn
  [z] = \left(z,\frac{z^2}{2},\ldots,\frac{z^n}{n},\ldots\right). 
\eeqnn
The bilinear equation (\ref{Psi-bilineq}) for the wave functions 
thereby turns into the bilinear equation 
\begin{multline}
  \oint\frac{dz}{2\pi i}z^{k+s'-s}e^{\xi(\bst'-\bst,z)/2}
    \tau(s',\bst'-[z^{-1}])\tau(s,\bst+[z^{-1}]) \\
= \oint\frac{dz}{2\pi i}z^{-k+s'-s}e^{\xi(\bst-\bst',z^{-1})/2}
    \tau(s'+1,\bst'+[z])\tau(s-1,\bst-[z]) 
\label{tau-bilineq}
\end{multline}
for the tau function, which holds for $k = 0,1,\ldots$ 
and arbitrary values of $s,s',\bst,\bst'$ 
under the condition (\ref{s's-condition}).  

Let us mention a few consequences of (\ref{tau-bilineq}).  
\begin{enumerate}
\item We can replace $z^{\pm k}$ 
by an arbitrary formal power series 
$f(z^{\pm 1}) = \sum_{k=0}^\infty f_kz^{\pm k}$ as
\begin{multline*}
  \oint\frac{dz}{2\pi i}f(z)z^{s'-s}e^{\xi(\bst'-\bst,z)/2}
    \tau(s',\bst'-[z^{-1}])\tau(s,\bst+[z^{-1}]) \\
= \oint\frac{dz}{2\pi i}f(z^{-1})z^{s'-s}e^{\xi(\bst-\bst',z^{-1})/2}
    \tau(s'+1,\bst'+[z])\tau(s-1,\bst-[z]). 
\end{multline*}
In particular, if we choose $f(z)$ as 
$f(z) = z^ke^{\xi(\bst'-\bst,z)/2}$, 
we have the bilinear equation 
\begin{multline}
  \oint\frac{dz}{2\pi i}z^{s'-s}e^{\xi(\bst'-\bst,z)}
    \tau(s',\bst'-[z^{-1}])\tau(s,\bst+[z^{-1}]) \\
= \oint\frac{dz}{2\pi i}z^{s'-s}
    \tau(s'+1,\bst'+[z])\tau(s-1,\bst-[z]), 
\label{tau-bilineq-alt}
\end{multline}
which is equivalent to (\ref{tau-bilineq}), hence 
may be thought of as yet another bilinear representation 
of the 1D Toda hierarchy.  
\item When $k = 0$ and $s' \ge s$, (\ref{tau-bilineq-alt}) 
reduces to the bilinear equation 
\beqnn
  \oint\frac{dz}{2\pi i}e^{\xi(\bst'-\bst,z)}
    \tau(s,\bst'-[z^{-1}])\tau(s,\bst+[z^{-1}]) = 0 
\eeqnn
of the KP ($s' = s$) or modified KP ($s' > s$) hierarchy. 
\item If we choose $s' = s$ and $\bst' = \bst$, 
(\ref{tau-bilineq}) reduces to 
\begin{multline*}
  \oint\frac{dz}{2\pi i}
    z^k\tau(s,\bst-[z^{-1}])\tau(s,\bst+[z^{-1}]) \\
= \oint\frac{dz}{2\pi i}
    z^{k-2}\tau(s+1,\bst+[z^{-1}])\tau(s-1,\bst-[z^{-1}]),
  \quad k = 0,1,2,\ldots. 
\end{multline*}
These equations imply that 
$\tau(s,\bst-[z^{-1}])\tau(s,\bst+[z^{-1}]) 
- z^{-2}\tau(s+1,\bst+[z^{-1}])\tau(s-1,\bst+[z^{-1}])$ 
is independent of $z$, hence a function of $s$ and $\bst$ only.  
Letting $z \to \infty$ shows that this function 
is equal to $\tau(s,\bst)^2$.  Thus we obtain 
the bilinear functional equation 
\begin{multline*}
  \tau(s,\bst-[z^{-1}])\tau(s,\bst+[z^{-1}]) \\
  = z^{-2}\tau(s+1,\bst+[z^{-1}])\tau(s-1,\bst-[z^{-1}])
    + \tau(s,\bst)^2
\end{multline*}
with a parameter $z$.  Expanded in powers of $z^{-1}$, 
the $z^{-2}$ part of this equation gives 
the Hirota equation 
\beqnn
  \frac{\rd^2\tau(s,\bst)}{\rd t_1^2}\tau(s,\bst) 
  - \left(\frac{\rd\tau(s,\bst)}{\rd t_1}\right)^2
  = \tau(s+1,\bst)\tau(s-1,\bst)  
\eeqnn
of the 1D Toda equation.  
\end{enumerate}

\subsection{Reduction from 2D Toda hierarchy}

The 2D Toda hierarchy has two sets of time variables 
$\bst = (t_1,t_2,\ldots)$ and 
$\bstbar = (\tbar_1,\tbar_2,\ldots)$.  
The Lax equations are formulated in terms 
of two Lax operators 
\beqnn
\begin{aligned}
  L &= e^{\rd_s} + u_1(s) + u_2(s)e^{-\rd_s} + \cdots,\\
  \Lbar &= \ubar_0(s)e^{\rd_s} + \ubar_1(s)e^{2\rd_s} + \cdots 
\end{aligned}
\eeqnn
and the generators of time evolutions 
\beqnn
  B_n = (L^n)_{\ge 0},\quad 
  \Bbar_n = (\Lbar^{-n})_{<0}
\eeqnn
as 
\beqnn
\begin{aligned}
  &\frac{\rd L}{\rd t_n} = [B_n,L],
  &\frac{\rd L}{\rd\tbar_n} = [\Bbar_n,L],
  &\quad\\
  &\frac{\rd\Lbar}{\rd t_n} = [B_n,\Lbar],
  &\frac{\rd\Lbar}{\rd\tbar_n} = [\Bbar_n,\Lbar],
  &\quad n = 1,2,\ldots.
\end{aligned}
\eeqnn
This hierarchy reduces to the 1D Toda hierarchy 
by adding the constraint
\footnote{Another reduction to the 1D Toda hierarchy 
is achieved by the constraint 
$L + L^{-1} = \Lbar + \Lbar^{-1}$ \cite{UT84}.  
This reduction is suited for the soliton solutions 
of the 1D Toda lattice.} 
\beqnn
  (\calL :=)\; L =\Lbar^{-1}. 
\eeqnn
Defining $\calL$ thus by both hand sides of this constraint 
and comparing the $(\quad)_{\ge 0}$ and $(\quad)_{<0}$ parts, 
we can readily see that $\calL$ can be written as 
\beqnn
  \calL = B_1 + C_1, 
\eeqnn
hence a difference operator of the form 
$e^{\rd_s} + b(s) + c(s)e^{-\rd_s}$.  
Moreover, under this constraint, we have the identities 
\beqnn
  B_n + \Bbar_n = \calL^n, 
\eeqnn
which imply that the time evolutions 
in the diagonal direction of the $(t_n,\tbar_n)$ plane 
are trivial: 
\beqnn
  \frac{\rd\calL}{\rd t_n} + \frac{\rd\calL}{\rd\tbar_n} 
  = [\calL^n, \calL] 
  = 0, 
  \quad n = 1,2,\ldots.  
\eeqnn
The residual time evolutions of $\calL$ generated by 
\beqnn
  A_n = \frac{1}{2}B_n - \frac{1}{2}\Bbar_n
\eeqnn
can be identified with the 1D Toda hierarchy.  

As regards the tau function, this reduction procedure 
amounts to adding the constraints 
\beqnn
  \frac{\rd\tau(s,\bst,\bstbar)}{\rd t_n}
  + \frac{\rd\tau(s,\bst,\bstbar)}{\rd\tbar_n} 
  = 0, \quad n = 1,2,\ldots
\eeqnn
to the tau function $\tau(\bst,\bstbar)$ 
of the 2D Toda hierarchy, which thereby becomes 
a function $\tau(s,\bst-\bstbar)$ of $s$ and $\bst-\bstbar$. 
The reduced function $\tau(s,\bst)$ is exactly 
the tau function of the 1D Toda hierarchy.

\section{$2+1$D extension}

\subsection{Lax equations and auxiliary linear equations}

Following Ogawa \cite{Ogawa08}, 
we introduce a new spatial variable $y$ 
and an infinite number of time variables 
$\bsx = (x_1,x_2,\ldots)$.  
The dynamical variables $b(s)$ and $c(s)$ 
now depend on $y,\bsx$ and $\bst$.  
The $2+1$D extension consists of the Toda flows 
with respect to $\bst$ and the commuting flows 
with respect to $\bsx$ defined by Lax equations 
of the form 
\beq
  \frac{\rd\calL}{\rd x_n} 
  = \calL^n\frac{\rd\calL}{\rd y} + [P_n,\calL] 
  = [\calL^n\rd_y + P_n, \calL], 
  \quad n = 1,2,\ldots,
\label{21-Lax-eq}
\eeq
where $\rd_y$ denotes $\rd/\rd y$, and 
$P_n$'s are difference operators of finite order 
specified below.  The associated auxiliary linear equations 
for $\Psi(s,z)$ and $\Psibar(s,z)$ read 
\beq
  \frac{\rd\Psi(s,z)}{\rd x_n} = (\calL^n\rd_ y + P_n)\Psi(s,z),\quad
  \frac{\rd\Psibar(s,z)}{\rd x_n} = (\calL^n\rd_y + P_n)\Psibar(s,z). 
\label{21-PsiP-lineq}
\eeq
The dressing operators $W$ and $\Wbar$ thereby 
satisfy the evolution equations 
\beq
  \frac{\rd W}{\rd x_n} = \calL^n\frac{\rd W}{\rd y} + P_nW,\quad
  \frac{\rd\Wbar}{\rd x_n} = \calL^n\frac{\rd\Wbar}{\rd y} + P_n\Wbar. 
\label{21-Sato-eq}
\eeq

$P_n$'s are determined by (\ref{21-Sato-eq}) themselves 
as follows.  Let us rewrite (\ref{21-Sato-eq}) as 
\beqnn
  P_n = \frac{\rd W}{\rd x_n}W^{-1} 
        - \calL^n\frac{\rd W}{\rd y}W^{-1}
      = \frac{\rd\Wbar}{\rd x_n}\Wbar^{-1} 
        - \calL^n\frac{\rd\Wbar}{\rd y}\Wbar^{-1}. 
\eeqnn
The $(\quad)_{\ge 0}$ and $(\quad)_{<0}$ parts 
of these equations give 
\beqnn
  (P_n)_{\ge 0} 
  = - \left(\calL^n\frac{\rd W}{\rd y}W^{-1}\right)_{\ge 0},\quad 
  (P_n)_{<0} 
  = - \left(\calL^n\frac{\rd\Wbar}{\rd y}\Wbar^{-1}\right)_{<0}. 
\eeqnn
Thus $P_n$'s are determined as 
\beq
  P_n 
  = - \left(\calL^n\frac{\rd W}{\rd y}W^{-1}\right)_{\ge 0} 
    - \left(\calL^n\frac{\rd\Wbar}{\rd y}\Wbar^{-1}\right)_{<0}.
\label{21-P}
\eeq

The auxiliary linear equations have another expression 
of the form 
\beq
  \frac{\rd\Psi(s,z)}{\rd x_n} = (z^n\rd_y + Q_n)\Psi(s,z),\quad
  \frac{\rd\Psibar(s,z)}{\rd x_n} = (z^{-n}\rd_y + Q_n)\Psibar(s,z),
\label{21-PsiQ-lineq}
\eeq
where 
\beq
  Q_n = P_n - \frac{\rd\calL^n}{\rd y} 
  = - \left(\frac{\rd W}{\rd y}e^{n\rd_s}W^{-1}\right)_{\ge 0} 
    - \left(\frac{\rd\Wbar}{\rd y}e^{-n\rd_s}\Wbar^{-1}\right)_{<0}. 
\label{21-Q}
\eeq

\subsection{Bilinear equation for wave functions}

Let us start from the bilinear equation 
\beqnn
  \oint\frac{dz}{2\pi i}
    z^k\Psi(s',\bsx,\bst',z)\Psi^*(s,\bsx,\bst,z) 
= \oint\frac{dz}{2\pi i}
    z^{-k}\Psibar(s',\bsx,\bst',z)\Psibar^*(s,\bsx,\bst,z) 
\eeqnn
of the 1D Toda hierarchy, and deform it 
to incorporate the auxiliary linear equations 
(\ref{21-PsiQ-lineq}).   To this end, 
we extend (\ref{21-PsiQ-lineq}) to higher orders as 
\beqnn
\begin{aligned}
  \prod_{i=1}^\infty 
  \left(\frac{\rd}{\rd x_i} - z^i\frac{\rd}{\rd y}
  \right)^{l_i}\Psi(s,z) 
  &= Q_{l_1,l_2,\ldots}\Psi(s,z),\\
  \prod_{i=1}^\infty 
  \left(\frac{\rd}{\rd x_i} - z^{-i}\frac{\rd}{\rd y}
  \right)^{l_i}\Psibar(s,z) 
  &= Q_{l_1,l_2,\ldots}\Psibar(s,z), 
\end{aligned}
\eeqnn
where $Q_{l_1,l_2,\ldots}$ are difference operators 
of finite order in $s$.  Applying  $Q_{l_1,l_2,\ldots}$ 
to both hand sides of the bilinear equation 
with respect to $s'$, we have the equations 
\begin{multline*}
  \oint\frac{dz}{2\pi i}
  z^k\prod_{i=1}^\infty 
  \left(\frac{\rd}{\rd x_i} - z^i\frac{\rd}{\rd y}
  \right)^{l_i}\Psi(s',\bsx,\bst',z)
  \cdot\Psi^*(s,\bsx,\bst,z)\\
= \oint\frac{dz}{2\pi i}
  z^{-k}\prod_{i=1}^\infty 
  \left(\frac{\rd}{\rd x_i} - z^{-i}\frac{\rd}{\rd y}
  \right)^{l_i}\Psibar(s',\bsx,\bst',z)
  \cdot\Psibar^*(s,\bsx,\bst,z)  
\end{multline*}
for all values of $l_1,l_2,\ldots$.  
These bilinear equations can be packed 
into the generating functional form 
\begin{multline}
  \oint\frac{dz}{2\pi i}
    z^k\Psi(s',y-\xi(\bsa,z),\bsx+\bsa,\bst',z)
    \Psi^*(s,y,\bsx,\bst,z)\\
= \oint\frac{dz}{2\pi i}
    z^{-k}\Psibar(s',y-\xi(\bsa,z^{-1}),\bsx+\bsa,\bst',z)
    \Psibar^*(s,y,\bsx,\bst,z) 
\label{21-Psi-bilineq}
\end{multline}
with new variables $\bsa = (a_1,a_2,\ldots)$.  
Note that this equation, like (\ref{Psi-bilineq}), 
holds for $k = 0,1,2,\ldots$ and 
arbitrary values of $s',s,\bsx,\bst',\bst$ 
under condition (\ref{s's-condition}).  

Moreover, we can extend (\ref{21-Psi-bilineq}) 
to the slightly more general (but actually equivalent) form 
\begin{multline}
  \oint\frac{dz}{2\pi i}
    z^k\Psi(s',y-\xi(\bsa,z),\bsx+\bsa,\bst',z)
    \Psi^*(s,y-\xi(\bsb,z),\bsx+\bsb,\bst,z)\\
= \oint\frac{dz}{2\pi i}
    z^{-k}\Psibar(s',y-\xi(\bsa,z^{-1}),\bsx+\bsa,\bst',z)
    \Psibar^*(s,y-\xi(\bsb,z^{-1}),\bsx+\bsb,\bst,z), 
\label{21-Psi-bilineqx}
\end{multline}
where $\bsb = (b_1,b_2,\ldots)$ is yet another set 
of variables.  This equation, too, 
holds for $k = 0,1,2,\ldots$ and arbitrary values 
of $s',s,\bsx,\bst',\bst'$ except for the condition 
(\ref{s's-condition}).  To derive this equation, 
we apply the operator $(-c\rd/\rd y)^l/l!$ 
(where $c$ is a constant and $l = 0,1,2,\ldots$) 
to both hand sides of (\ref{21-Psi-bilineq}), 
shift $k$ to $k+ln$ ($n = 1,2,\ldots$), 
and take the summation over $l = 0,1,2,\ldots$.  
The outcome is the equation 
\begin{multline*}
  \oint\frac{dz}{2\pi i}
    z^k\Psi(s',y-\xi(\bsa,z)-cz^n,\bsx+\bsa,\bst',z)
    \Psi^*(s,y-cz^n,\bsx,\bst,z)\\
= \oint\frac{dz}{2\pi i}
    z^{-k}\Psibar(s',y-\xi(\bsa,z)-cz^{-n},\bsx+\bsa,\bst',z)
    \Psibar^*(s,y-cz^{-n},\bsx,\bst,z).
\end{multline*}
Repeating this procedure for $n = 1,2,\ldots$ 
with independent constants $c = b_n$, 
we can derive the equation 
\begin{multline*}
  \oint\frac{dz}{2\pi i}
    z^k\Psi(s',y-\xi(\bsa+\bsb,z),\bsx+\bsa,\bst',z)
    \Psi^*(s,y-\xi(\bsb,z),\bsx,\bst,z)\\
= \oint\frac{dz}{2\pi i}
    z^{-k}\Psibar(s',y-\xi(\bsa+\bsb,z),\bsx+\bsa,\bst',z)
    \Psibar^*(s,y-\xi(\bsb,z^{-1}),\bsx,\bst,z). 
\end{multline*}
Replacing $\bsx \to \bsx - \bsb$ and $\bsa \to \bsa - \bsb$ 
in this equation, we obtain (\ref{21-Psi-bilineqx}).

\subsection{Bilinear equation for tau function}

Let $\tau(s,\bsx,\bst)$ be a tau function 
in the sense of the 1D Toda hierarchy, namely, 
a function with which the wave functions 
are expressed as (\ref{Psi-tau}).  Note that 
such a tau function is unique up to a multiplier 
that depends on only $\bsx$. 

The bilinear equation (\ref{21-Psi-bilineq}) 
for the wave functions turns into an equation 
for the tau function of the form 
\begin{multline*}
\begin{aligned}
 &\oint\frac{dz}{2\pi i}z^{k+s'-s}e^{\xi(\bst'-\bst,z)/2}\\
 &\quad \times
    \frac{\tau(s',y-\xi(\bsa,z),\bsx+\bsa,\bst'-[z^{-1}])
          \tau(s,y,\bsx,\bst+[z^{-1}])}
    {\tau(s',y-\xi(\bsa,z),\bsx+\bsa,\bst')\tau(s,y,\bsx,\bst)}
\end{aligned}\\
\begin{aligned}
=&\oint\frac{dz}{2\pi i}z^{-k+s'-s}e^{\xi(\bst-\bst',z^{-1})/2}\\
 &\quad \times
    \frac{\tau(s'+1,y-\xi(\bsa,z^{-1}),\bsx+\bsa,\bst'+[z])
          \tau(s-1,y,\bsx,\bst-[z])}
    {\tau(s',y-\xi(\bsa,z^{-1}),\bsx+\bsa,\bst')\tau(s,y,\bsx,\bst)}. 
\end{aligned}
\end{multline*}
We can now use the same trick as used in Section 2.4. 
Namely, we can replace $z^{\pm k}$ by an arbitrary power series
$f(z^{\pm 1}) = \sum_{k=0}^\infty f_kz^{\pm k}$ of $z$ as
\begin{multline*}
\begin{aligned}
 &\oint\frac{dz}{2\pi i}f(z)z^{s'-s}e^{\xi(\bst'-\bst,z)/2}\\
 &\quad \times
    \frac{\tau(s',y-\xi(\bsa,z),\bsx+\bsa,\bst'-[z^{-1}])
          \tau(s,y,\bsx,\bst+[z^{-1}])}
    {\tau(s',y-\xi(\bsa,z),\bsx+\bsa,\bst')\tau(s,y,\bsx,\bst)}
\end{aligned}\\
\begin{aligned}
=&\oint\frac{dz}{2\pi i}f(z^{-1})z^{s'-s}e^{\xi(\bst-\bst',z^{-1})/2}\\
 &\quad \times
    \frac{\tau(s'+1,y-\xi(\bsa,z^{-1}),\bsx+\bsa,\bst'+[z])
          \tau(s-1,y,\bsx,\bst-[z])}
    {\tau(s',y-\xi(\bsa,z^{-1}),\bsx+\bsa,\bst')\tau(s,y,\bsx,\bst)}. 
\end{aligned}
\end{multline*}
In particular, if we choose $f(z)$ as 
\beqnn
  f(z) = z^k\tau(s',y-\xi(\bsa,z),\bsx+\bsa,\bst')\tau(s,y,\bsx,\bst), 
\eeqnn
the denominators disappear and we obtain 
the bilinear equation 
\begin{multline}
\begin{aligned}
 &\oint\frac{dz}{2\pi i}z^{k+s'-s}e^{\xi(\bst'-\bst,z)/2}\\
 &\quad \times
    \tau(s',y-\xi(\bsa,z),\bsx+\bsa,\bst'-[z^{-1}])
    \tau(s,y,\bsx,\bst+[z^{-1}]) 
\end{aligned} \\
\begin{aligned}
=&\oint\frac{dz}{2\pi i}z^{-k+s'-s}e^{\xi(\bst-\bst',z^{-1})/2}\\
 &\quad \times
    \tau(s'+1,y-\xi(\bsa,z^{-1}),\bsx+\bsa,\bst'+[z])
    \tau(s-1,y,\bsx,\bst-[z])
\end{aligned}
\label{21-tau-bilineq}
\end{multline}
for the tau function.  

In the same way, the bilinear equation (\ref{21-Psi-bilineqx}) 
of a slightly more general form can be converted to 
\begin{multline}
\begin{aligned}
 &\oint\frac{dz}{2\pi i}z^{k+s'-s}e^{\xi(\bst'-\bst,z)/2}
    \tau(s',y-\xi(\bsa,z),\bsx+\bsa,\bst'-[z^{-1}])\\
 &\quad \times
    \tau(s,y-\xi(\bsb,z),\bsx+\bsb,\bst+[z^{-1}]) 
\end{aligned}\\
\begin{aligned}
=&\oint\frac{dz}{2\pi i}z^{-k+s'-s}e^{\xi(\bst-\bst',z^{-1})/2}
    \tau(s'+1,y-\xi(\bsa,z^{-1}),\bsx+\bsa,\bst'+[z])\\
 &\quad \times
    \tau(s-1,y-\xi(\bsb,z^{-1}),\bsx+\bsb,\bst-[z]). 
\end{aligned}
\label{21-tau-bilineqx}
\end{multline}

\section{Logarithmic extension}

\subsection{Lax equations}

Following Carlet, Dubrovin and Zhang \cite{CDZ04}, 
we define the logarithm $\log\calL$ 
of the Lax operator $\calL$ as 
\beqnn
  \log\calL = \frac{1}{2}W\rd_sW^{-1} - \frac{1}{2}\Wbar\rd_s\Wbar^{-1}. 
\eeqnn
This definition can be rewritten as 
\beqnn
  \log\calL 
  = - \frac{1}{2}[\rd_s,W]W^{-1} + \frac{1}{2}[\rd_s,\Wbar]\Wbar^{-1} 
  = - \frac{1}{2}\frac{\rd W}{\rd s}W^{-1} 
    + \frac{1}{2}\frac{\rd\Wbar}{\rd s}\Wbar^{-1}, 
\eeqnn
which shows that $\log\calL$ becomes a difference operator 
(of infinite order).  

The logarithmic extension of the Toda hierarchy 
consists of the Toda flows with respect to $\bst$ 
and another set of commuting flows with respect to 
$\bsx = (x_1,x_2,\ldots)$.  The extended flows 
are defined by the Lax equations \cite{CDZ04} 
\beq
  \frac{\rd\calL}{\rd x_n} = [C_n,\calL], 
  \quad n = 1,2,\ldots,
\label{log-Lax-eq}
\eeq
where 
\beqnn
  C_n = \left(\calL^n\log\calL\right)_{\ge 0} 
        - \left(\calL^n\log\calL\right)_{<0}. 
\eeqnn
Note that $\calL^n\log\calL$ can be expressed 
in terms of the dressing  operators as 
\beq
  \calL^n\log\calL 
  = \frac{1}{2}We^{n\rd_s}\rd_sW^{-1} 
  - \frac{1}{2}\Wbar e^{-n\rd_s}\rd_s\Wbar^{-1}. 
\label{L^nlogL}
\eeq

A few remarks are in order.   
\begin{enumerate}
\item This definition of $C_n$'s differs 
from the usual definition 
\beqnn
  C_n = \left(\calL^n(\log\calL - c_n)\right)_{\ge 0} 
        - \left(\calL^n(\log\calL - c_n)\right)_{<0}, 
\eeqnn
where $c_n$'s are numerical constants of the form 
\beqnn
  c_n = 1 + 2 + \cdots + \frac{1}{n} 
\eeqnn
that plays an important role in the application 
to 2D topological field theories \cite{EY94,Getzler01,Zhang02}. 
In the context of integrable structure, however, 
this difference is superficial.  
\item Since $C_n$ can be expressed as 
\beqnn
  C_n = 2\left(\calL^n\log\calL\right)_{\ge 0} - \calL^n\log\calL 
      = - 2\left(\calL^n\log\calL\right)_{<0} + \calL^n\log\calL 
\eeqnn
and $\calL^n\log\calL$ commutes with $\calL$, 
we can rewrite the Lax equations as 
\beq
  \frac{\rd\calL}{\rd x_n} 
  = [2\left(\calL^n\log\calL\right)_{\ge 0},\calL] 
  = [-2\left(\calL^n\log\calL\right)_{<0},\calL]. 
\label{log-Lax-eqx}
\eeq
\end{enumerate}

\subsection{Auxiliary linear equations}

For comparison with the $2+1$D extension, 
let us rewrite the Lax equations (\ref{log-Lax-eqx}).  
Note that $2\left(\calL^n\log\calL\right)_{\ge 0}$ 
can be expressed as 
\beqnn
\begin{aligned}
  2\left(\calL^n\log\calL\right)_{\ge 0}
  &= - \left(\calL^n\frac{\rd W}{\rd s}W^{-1}\right)_{\ge 0} 
     + \left(\calL^n\frac{\rd\Wbar}{\rd s}\Wbar^{-1}\right)_{\ge 0}\\
  &= P_n + \calL^n\frac{\rd\Wbar}{\rd s}\Wbar^{-1}, 
\end{aligned}
\eeqnn
where 
\beq
  P_n 
  = - \left(\calL^n\frac{\rd W}{\rd s}W^{-1}\right)_{\ge 0} 
    - \left(\calL^n\frac{\rd\Wbar}{\rd s}\Wbar^{-1}\right)_{<0}. 
\label{log-P}
\eeq
We can further rewrite the right hand side as 
\beqnn
\begin{aligned}
  2\left(\calL^n\log\calL\right)_{\ge 0}
  &= P_n + \calL^n[\rd_s,\Wbar]\Wbar^{-1}\\
  &= P_n + \calL^n\rd_s - \Wbar e^{-n\rd_s}\rd_s\Wbar^{-1}.  
\end{aligned}
\eeqnn
Since the last term $\Wbar e^{-n\rd_s}\rd_s\Wbar^{-1}$ 
commutes with $\calL$, we can remove it and obtain 
the equations 
\beq
  \frac{\rd\calL}{\rd x_n} = [\calL^n\rd_s + P_n, \calL]. 
\label{log-LaxP-eq}
\eeq

Written in this form, the Lax equations 
of the logarithmic extension exhibit 
remarkable similarity with the Lax equations 
(\ref{21-Lax-eq}) of the $2+1$D extensions. 
The only difference is that the role of $y$ 
is now played by $s$.  Thus the logarithmic extension 
may be thought of as a kind of dimensional reduction 
(identifying $\rd_y$ with $\rd_s$) of the $2+1$D extension.  
Inspired by this observation, we can readily find 
the evolution equations 
\beq
  \frac{\rd W}{\rd x_n} 
  = \calL^n\frac{\rd W}{\rd s} + P_nW,\quad 
  \frac{\rd\Wbar}{\rd x_n} 
  = \calL^n\frac{\rd\Wbar}{\rd s} + P_n\Wbar 
\label{log-Sato-eq}
\eeq
for the dressing operators as counterparts of 
(\ref{21-Sato-eq}).  

This is, however, a place where a significant difference 
also shows up.  In the present case, we can further 
rewrite (\ref{log-Sato-eq}) to such a form as 
\beq
\begin{aligned}
  \frac{\rd W}{\rd x_n} 
  &= (\calL^n\rd_s + P_n)W - We^{n\rd_s}\rd_s,\\
  \frac{\rd\Wbar}{\rd x_n} 
  &= (\calL^n\rd_s + P_n)\Wbar - \Wbar e^{-n\rd_s}\rd_s, 
\end{aligned}
\label{log-Sato-eqx}
\eeq
which rather resembles (\ref{Sato-eq}).  
Note here that the roles of $e^{\pm n\rd_s}/2$ 
in (\ref{Sato-eq}) are now played by $e^{\pm n\rd_s}\rd_s$, 
which are connected with $\calL^n\log\calL$ 
by the dressing operators as shown in (\ref{L^nlogL}).  
These ``undressed'' generators of time evolutions 
determine the exponential factors of the wave functions.  
The exponential factors $e^{\xi(\bst,z^{\pm 1})/2}$ 
are thus generated from $z^s$ by the first set of 
generators $e^{\pm n\rd_s}/2$ as 
\beqnn
  \exp\left(\sum_{n=1}^\infty t_ne^{\pm n\rd_s}/2\right)z^s 
  = z^se^{\xi(\bst,z^{\pm 1})/2}. 
\eeqnn
In the same sense, the second set of generators 
$e^{\pm n\rd_s}\rd_s$ give the power (rather than 
exponential) functions $z^{\xi(\bsx,z^{\pm 1})}$ as 
\beqnn
  \exp\left(\sum_{n=1}^\infty x_ne^{\pm n\rd_s}\rd_s\right)z^s 
  = z^{s+\xi(\bsx,z^{\pm 1})}.
\eeqnn

Bearing the last observation in mind, 
we introduce the wave functions 
\beqnn
  \Psi(s,z) = Wz^{s+\xi(\bsx,z)}e^{\xi(\bst,z)/2}
  = \left(1 + \sum_{n=1}^\infty w_n(s)z^{-n}\right) 
    z^{s+\xi(\bsx,z)}e^{\xi(\bst,z)/2},\\
  \Psibar(s,z) = \Wbar z^{s+\xi(\bsx,z^{-1})}e^{-\xi(\bst,z^{-1})/2} 
  = \left(\sum_{n=0}^\infty \wbar_n(s)z^n\right)
    z^{s+\xi(\bsx,z^{-1})}e^{-\xi(\bst,z^{-1})/2}. 
\eeqnn
(\ref{log-Sato-eqx}) can be thereby converted to 
the auxiliary linear equations 
\beq
  \frac{\rd\Psi(s,z)}{\rd x_n} = (\calL^n\rd_s + P_n)\Psi(s,z),\quad
  \frac{\rd\Psibar(s,z)}{\rd x_n} = (\calL^n\rd_s + P_n)\Psibar(s,z). 
\label{log-PsiP-lineq}
\eeq
As in the case of the $2+1$-dimensional extension, 
these auxiliary linear equations have another expression 
of the form 
\beq
  \frac{\rd\Psi(s,z)}{\rd x_n} = (z^n\rd_s + Q_n)\Psi(s,z),\quad 
  \frac{\rd\Psibar(s,z)}{\rd x_n} = (z^{-n}\rd_s + Q_n)\Psibar(s,z), 
\label{log-PsiQ-lineq}
\eeq
where 
\beq
  Q_n = P_n - \frac{\rd\calL^n}{\rd s}
  = - \left(\frac{\rd W}{\rd s}e^{n\rd_s}W^{-1}\right)_{\ge 0} 
    - \left(\frac{\rd\Wbar}{\rd s}e^{-n\rd_s}\Wbar^{-1}\right)_{<0}. 
\label{log-Q}
\eeq

\subsection{Bilinear equations}

Since the structure of the auxiliary linear equations 
(\ref{log-PsiQ-lineq}) is almost the same as 
those of the $2+1$D extension, 
we can convert these auxiliary linear equations 
into a bilinear form in exactly the same way.  
Thus, defining the dual wave functions as 
\beqnn
  \Psi^*(s,z) = V^*z^{-s-\xi(\bsx,z)}e^{-\xi(\bst,z)/2},\quad 
  \Psibar^*(s,z) = \Vbar^*z^{-s-\xi(\bsx,z^{-1})}e^{-\xi(\bst,z^{-1})/2}, 
\eeqnn
we obtain the bilinear equation
\begin{multline}
  \oint\frac{dz}{2\pi i}
    z^k\Psi(s'-\xi(\bsa,z),\bsx+\bsa,\bst',z)
    \Psi^*(s-\xi(\bsb,z),\bsx+\bsb,\bst,z)\\
= \oint\frac{dz}{2\pi i}
    z^{-k}\Psibar(s'-\xi(\bsa,z^{-1}),\bsx+\bsa,\bst',z)
    \Psibar^*(s-\xi(\bsb,z^{-1}),\bsx+\bsb,\bst,z), 
\label{log-Psi-bilineqx}
\end{multline}
which holds for $k = 0,1,2,\ldots$ and 
arbitrary values of $s',s,\bsx,\bst',\bst$ 
except for the condition (\ref{s's-condition}).  

It deserves to be stressed here that the integrands 
in the contour integrals are single-valued.  
The multi-valuedness of the power functions 
$z^{s+\xi(\bsx,z^{\pm 1})}$ in the wave functions 
and the dual wave functions cancels each other. 
This cancellation mechanism is based on the special shift 
\beqnn
  s' \to s' - \xi(\bsa,z^{\pm 1}),\quad \bsx \to \bsx + \bsa,\quad
  s \to s - \xi(\bsb,z^{\pm 1}),\quad \bsx \to \bsx + \bsb 
\eeqnn
of the $s$ and $\bsx$ variables in the integrand.  
Actually, this special shift was a main mystery 
of Milanov's bilinear formalism;  
we can now explain its origin in the $2+1$D extension. 

Lastly, by the same trick as used in the derivation 
of (\ref{21-tau-bilineq}) and (\ref{21-tau-bilineqx}), 
we can derive from (\ref{log-Psi-bilineqx}) 
the bilinear equation 
\begin{multline}
\begin{aligned}
 &\oint\frac{dz}{2\pi i}z^{k+s'-s}e^{\xi(\bst'-\bst,z)/2}
    \tau(s'-\xi(\bsa,z),\bsx+\bsa,\bst'-[z^{-1}])\\
 &\quad \times
    \tau(s-\xi(\bsb,z),\bsx+\bsb,\bst+[z^{-1}]),
\end{aligned}\\
\begin{aligned}
=&\oint\frac{dz}{2\pi i}z^{-k+s'-s}e^{\xi(\bst-\bst',z^{-1})/2}
    \tau(s'+1-\xi(\bsa,z^{-1}),\bsx+\bsa,\bst'+[z])\\
 &\quad \times
    \tau(s-1-\xi(\bsb,z^{-1}),\bsx+\bsb,\bst-[z]) 
\end{aligned}
\label{log-tau-bilineqx}
\end{multline}
for the tau function.  This equation contains 
Milanov's bilinear equation as a special case.

\section{Conclusion}

We have thus shown that the $2+1$D extension 
and the logarithmic extension have a quite parallel 
structure.  Relevant equations of these two extended 
Toda hierarchy can be paired as follows: 
\begin{itemize}
\item Lax equations: 
(\ref{21-Lax-eq}) $\leftrightarrow$ (\ref{log-Lax-eq})
\item Auxiliary linear equations: 
(\ref{21-PsiP-lineq}), (\ref{21-P}) 
  $\leftrightarrow$ (\ref{log-PsiP-lineq}), (\ref{log-P}) 
\item Evolution equations of dressing operators: 
(\ref{21-Sato-eq}) $\leftrightarrow$ (\ref{log-Sato-eq})
\item Another form of auxiliary linear equations: 
(\ref{21-PsiQ-lineq}), (\ref{21-Q}) 
  $\leftrightarrow$ (\ref{log-PsiQ-lineq}), (\ref{log-Q})
\item Bilinear equations of wave functions: 
(\ref{21-Psi-bilineqx}) $\leftrightarrow$ (\ref{log-Psi-bilineqx})
\item Bilinear equations of tau functions: 
(\ref{21-tau-bilineqx}) $\leftrightarrow$ (\ref{log-tau-bilineqx})
\end{itemize}
A new feature of the logarithmic extension 
is the emergence of the multi-valued factor 
$z^{\xi(\bsx,z^{\pm 1})}$ in the wave functions.  
The multi-valuedness, however, disappears 
in the integrand of the bilinear equations.  
This fact plays a role in the heuristic part of 
Milanov's derivation of bilinear equations \cite{Milanov05}.  
In our approach, this cancellation mechanics 
of multi-valuedness is rather a consequence of 
dimensional reduction of the $2+1$D extension.  

Our approach can be readily generalized to 
the reduction  of the 2D Toda hierarchy defined 
by the constraint
\beqnn
  (\calL :=)\;   L^{N} = \Lbar^{-\Nbar}, 
\eeqnn
where $N$ and $\Nbar$ are arbitrary positive integers.  
The reduced Lax operator $\calL$ thus defined 
takes such a form as 
\beqnn
  \calL = B_N + \Bbar_{\Nbar} 
  = e^{N\rd_s} + b_1(s)e^{(N-1)\rd_s} + b_N(s) 
     + c_1(s)e^{-\rd_s} + \cdots + c_{\Nbar}(s)e^{-\Nbar\rd_s}. 
\eeqnn
The logarithmic extension of this reduced hierarchy 
coincides with Carlet's ``extended bigraded 
Toda hierarchy'' \cite{Carlet06}.  We can derive 
bilinear equations for the wave functions 
and the tau functions, which contains bilinear equations 
derived by Milanov and Tseng \cite{MT08} as a special case.

\subsection*{Acknowledgements}
The author thanks Saburo Kakei 
for useful comments and discussion.  
This work is partly supported by Grant-in-Aid for 
Scientific Research No. 19104002, No. 19540179 and No. 21540218 
from the Japan Society for the Promotion of Science.

\end{document}